\documentclass[aps,prb,twocolumn,groupedaddress]{revtex4-1}

\usepackage{graphicx}
\usepackage[usenames]{color}
\usepackage{siunitx}

\begin{document}

\title{Charge transfer from an adsorbed ruthenium-based photosensitizer through an ultra-thin aluminium oxide layer and into a metallic substrate}

\author{Andrew J. Gibson}
\author{Robert H. Temperton}
\author{Karsten Handrup}
\author{Matthew Weston}
\author{Louise C. Mayor}
\author{James N. O'Shea}
\email{james.oshea@nottingham.ac.uk}
\affiliation{School of Physics and Astronomy \& Nottingham
Nanotechnology and Nanoscience Centre (NNNC), University of
Nottingham, Nottingham, NG7 2RD, U.K.}

\begin{abstract}
The interaction of the dye molecule N3 (cis-bis(isothiocyanato)bis(2,2-bipyridyl-4,4'-dicarboxylato)-ruthenium(II)) with the ultra-thin oxide layer on a AlNi(110) substrate, has been studied using synchrotron radiation based photoelectron spectroscopy, resonant photoemission spectroscopy (RPES) and near edge X-ray absorption fine structure spectroscopy (NEXAFS). Calibrated X-ray absorption and valence band spectra of the monolayer and multilayer coverages reveal that charge transfer is possible from the molecule to the AlNi(110) substrate via tunnelling through the ultra-thin oxide layer and into the conduction band edge of the substrate. This charge transfer mechanism is possible from the LUMO+2\&3 in the excited state but not from the LUMO, therefore enabling core-hole clock analysis, which gives an upper limit of $6.0\pm$\SI{2.5}{fs} for the transfer time. This indicates that ultra-thin oxide layers are a viable material for use in dye-sensitized solar cells (DSSC), which may lead to reduced recombination effects and improved efficiencies of future devices.

\end{abstract}

\maketitle
\section{Introduction}

In a dye-sensitised solar cell (DSSC), a photo excited dye injects an electron into the conduction band of the substrate. DSSCs offer a potential lower cost alternative to their inorganic counterparts, with cheaper materials and manufacturing processes.\cite{Green2004}  They can also be engineered into flexible sheets, making them easier to transport and install.\cite{Razykov2011} DSSC are currently less efficient than other thin film technologies such as CuInxGa1-xSe2 or CdTe.\cite{NREL} However, the lower cost makes them a very competitive alternative. Ultra-thin oxide layers have already improved the performance of inorganic solar cells, by reducing recombination affects at surface defects through passivation.\cite{Dingemans2012}  In DSSCs reduced recombination rates have been attributed to surface passivation, with the aluminium oxide coating serving as a tunneling barrier between the redox mediator and conductive electrons.\cite{Pascoe2013} Hence aluminium oxide layers between a dye and TiO$_{2}$ substrate have been shown to increase power conversion efficiency,\cite{Kim2010a} by reducing dark current and increasing electron lifetimes.\cite{Roelofs2013} An optimal oxide layer of $14\pm2 \dot A$ has been shown to improve efficiency via the reduction recombination rates.\cite{Neo2011} Since an aluminium oxide layer suppresses the injection of an electron into the substrate,\cite{Makinen2011} an ultra-thin oxide may provide an attractive option for future DSSC devices by passivation of the surface, therefore reducing recombination effect with minimal disruption to charge injection. However the potential for ultra-thin oxide layers to be used for this application has received little attention. In this paper we investigate charge transfer from a dye molecule through an ultra-thin oxide layer and into the underlying substrate, thus giving important information on the fundamental processes which will affect future DSSC devices.

\begin{figure}[!h]
\centering
\includegraphics[width=8cm]{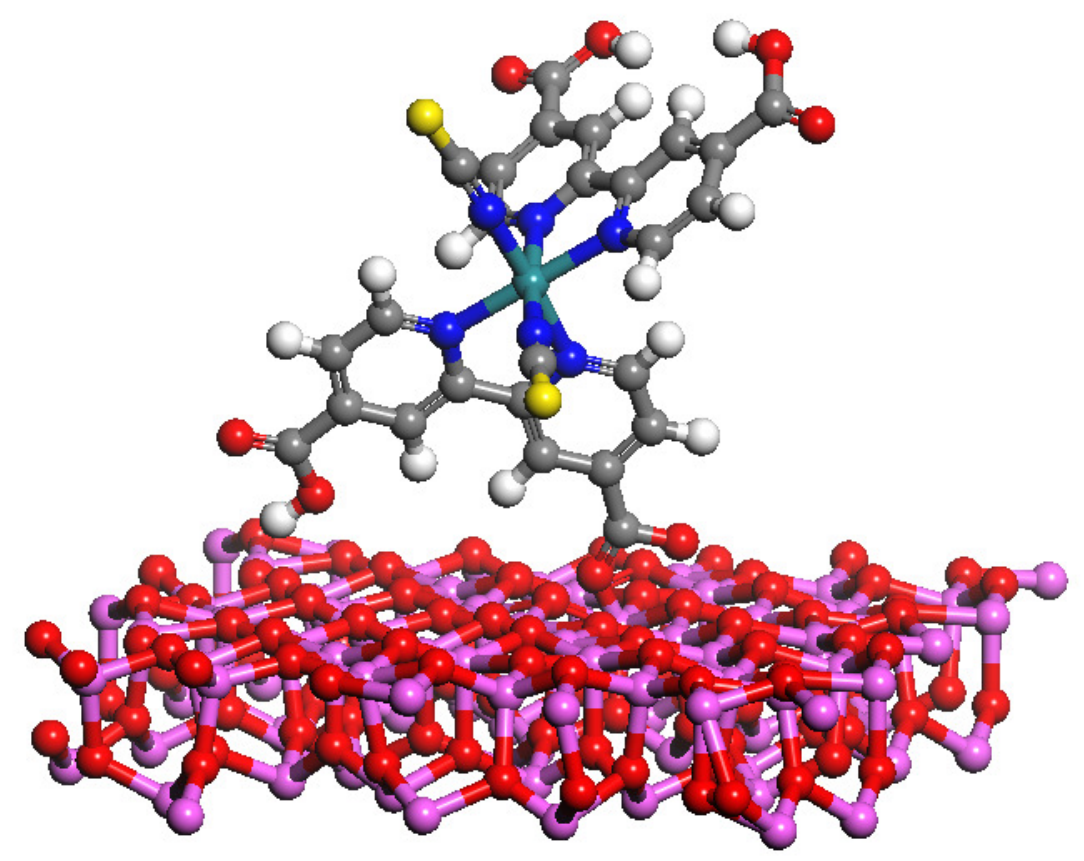}
\caption{Schematic of N3 molecule on ultra-thin aluminium oxide layer. H - White, C - Grey, N - Blue, S - Yellow, Al - Purple. The underlying AlNi(110) substrate is omitted.}
\label{fig:N3_molecule}
\end{figure}

The dye molecule N3 (cis-bis(isothiocyanato)bis(2,2-bipyridyl-4,4'-dicarboxylato)-ruthenium(II)), shown in FIG. \ref{fig:N3_molecule}, was chosen because of its performance as a photosensitive molecule used in the construction of DSSCs. It was unmatched for 8 years and as such it has become the benchmark for heterogeneous charge transfer in mesoporous solar cells.\cite{Gratzel2004} For N3, if rapid charge injection can take place out of the molecule then chemical transformations cannot occur. This leads to a highly stable solar cell arrangement that can operate for approximately \SI{20}{years}, without noticeable loss in performance.\cite{Gratzel2004}

The aluminium oxide surface is one of the most intensely studied metal oxide surfaces, but most previous applications relate to models of catalysts.\cite{Diebold2010} Its structure is similar to bulk Al$_{2}$O$_{3}$, hence the film is an insulator with a band gap of \SI{6.7}{eV}.\cite{Andersson1999}  A well ordered oxide layer is not formed on pure aluminium,\cite{Trost1998} however, oxidation of clean AlNi(110) does form a well ordered, self terminating, ultra-thin oxide layer on the surface. The stoichiometry of this oxide is Al$_{10}$O$_{13}$,\cite{Kresse2005} and previous studies have reported on the spectroscopy and organisation of this surface.\cite{Martin2011,Kresse2005} The film self terminates at three atomic layers because oxygen in the gas phase does not dissociate on the oxygen terminated surface.\cite{Diebold2010}  Since the oxide layer is only three atomic layers thick, there is potential for electrons to easily tunnel from the dye through the oxide and into the underlying AlNi substrate, while still reducing recombination affects. Thus Al$_{10}$O$_{13}$ on AlNi(110) provides a model for a DSSC device which could be developed in the future.

In order to develop our understanding of N3 on Al$_{10}$O$_{13}$, we must first consider its bonding. Prior research of N3 on Au(111) the bonding is dominated by the sulphur atom on the thiocyanate and van der Waals forces from the remainder of the molecule.\cite{Mayor2009a}  On TiO$_{2}$ N3 has shown bonding via de-protonation of two carboxylic acid groups on the bi-isonicotinic acid ligand, and via the interaction of sulphur on the thiocyanate ligand.\cite{Mayor2008a} Since the surface in this experiment is oxygen terminated it is expected that N3 will form chemical bonds with the surface in a manner similar to that seen on TiO$_{2}$.\cite{Mayor2008a} We have also previously reported the charge transfer interaction of N3 and its bi-isonicotinic acid ligand on Au(111)\cite{Britton2011,Taylor2007a} and N3 as well as related water splitting molecules on TiO$_{2}$.\cite{Mayor2008a,Weston2011a,Taylor2007} The upper limit for the charge transfer time from N3 to Au(111) and TiO$_{2}$ was \SI{4.4}{fs}\cite{Britton2011} and \SI{12}{fs}\cite{Weston2011a} respectively.

In order to further develop DSSCs it's important that we understand the subtle bonding and electronic properties that lead to efficient photon to current efficiencies. Here we present X-ray photoemission spectroscopy (XPS) results that allow characterisation of bonding between the molecule and the surface. Resonant photoemission spectroscopy (RPES) as well as near edge X-ray absorption fine structure (NEXAFS) that allow measurement of charge transfer interactions between the surface and the molecule.

\section{Experiment}

Experiments were carried out at the D1011 bending magnetic beamline at the MAX-lab, Swedish synchrotron radiation facility. The beamline covers photon energies in the range 30 to \SI{1600} {eV}, the end station is equipped with a SCIENTA SES200 (upgraded) electron energy analyzer and an MCP detector for electron yield measurements. The baseline operating pressure was $3\times10^{-10}$ mbar.

The sample used was a single crystal AlNi(110) which was sputtered at 1kV and then flash annealed to \SI{1300}{K} via ebeam heating. Subsequent sputtering and annealing cycles were repeated until C 1\textit{s} and O 1\textit{s} peaks were no longer observed in XPS. The oxide layer was built by dosing the sample with \SI{1800}{L} of O$_{2}$ at $3\times10^{-6}$ mbar and \SI{600}{K}, followed by an anneal \SI{900}{K} for \SI{10}{mins}. Further details on the formation and detailed description of this oxide layer are given elsewhere.\cite{Jaeger1991,Martin2011} Other papers\cite{Heinke2010,Schmid2006} have recommended a two step oxidation process, with high temperature (\SI{1050}{K}), in order to close open metal patches in the oxide layer. However during our analysis of the surface it became clear that slightly higher temperature annealing caused a significant reduction in the oxide signal in XPS. We therefore adopted a lower temperature anneal following oxidation.

N3 obtained from Solaronix SA, Switzerland, was deposited via a ultra-high vacuum compatible electrospray deposition system (MolecularSpray, UK) with methodology described elsewhere.\cite{Mayor2008a} N3 has been shown to remain stable after electrospray deposition onto TiO$_2$\cite{Mayor2008a,Weston2011a,Kley2014} and Au(111).\cite{Mayor2009a,Britton2011} The dye molecule was dissolved in a solution of 3(methanol):1(water) and sprayed for \SI{90}{mins} during which the pressure rose to $2\times10^{-7}$ mbar due to gas load from the electrospray system and the presence of solvent molecules in the molecule beam. This formed a deposition spot on the surface a few millimeters in diameter with a range of coverages in a Gaussian distribution, from partial monolayer at the edges to multilayer in the centre. The multilayer data was combined with data obtained in previously published experiments of N3.\cite{Mayor2008a,Weston2011a}

All measurements were performed at room temperature. XPS data were calibrated to the Fermi edge. A Shirley background\cite{Shirley1972} was removed and the spectra normalised to the photon flux and the number of sweeps, before curve-fit analysis using pseudo-Voigt functions.\cite{Kielkopf1973} NEXAFS data were recorded at the N 1\textit{s} adsorption edge with the emitted electrons collected by a partial yield detector with a retardation potential of \SI{200}{V}. For NEXAFS and RPES the photon energy was calibrated by taking the energy separation of the Al 2\textit{p} core-level photoemission peaks excited by X-rays in first and second order.

To calculate coverages the inelastic mean free path (IMFP) $\lambda$ (in \AA) through the molecule was approximated as a function of electron energy and calculated via the TPP-2M predictive equation.\cite{Tanuma2003} This was then used to calculate surface concentration by looking at the increased intensity of the adsorbate C 1\textit{s} peak in comparison to the suppression of the Al 2\textit{p} surface. The IMFP was used in the Carley-Roberts formula\cite{Carley1978} to calculate surface concentration, coverage was then calculated via the footprint of the molecule as measured in DFT analysis. In the multilayer the number of photoelectrons in PES, which came from substrate, or the layer of molecules, which were bonded to the substrate, was negligible. Hence this represents molecules that are isolated from the surface. 

\section{Results and Discussion}
\subsection{Adsorption bonding}
\begin{figure}[!h]
\centering
\includegraphics[width=8cm]{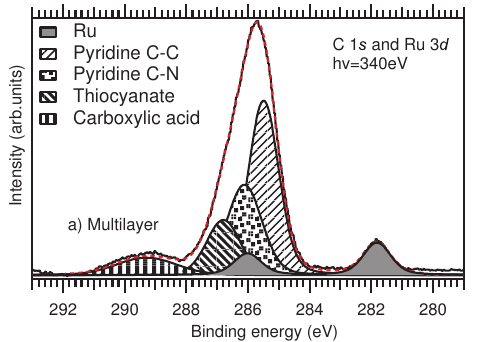}
\includegraphics[width=8cm]{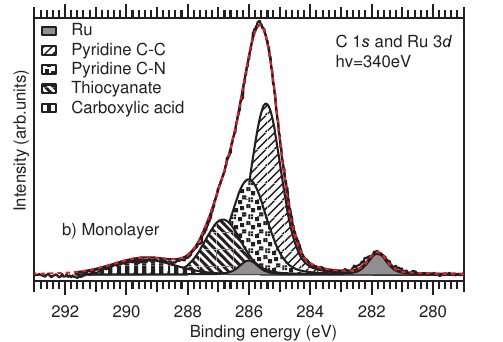}
\caption{C 1\textit{s} plot fit for (a) multilayer data and (b) monolayer, with different peaks for each of the different bonding environments of carbon. The ratios of the peaks fit with expected result for the intact molecule. Peak positions are the same across all coverages with the exception of the carboxylic acid peak, which is involved in intermolecular bonding in monolayer and multilayer coverages. h$\nu$=\SI{340}{eV}.}
\label{N3_C1s_multi_mono_fit}
\end{figure}

To build a complete picture of the interaction between N3 and Al$_{10}$O$_{13}$ we must first consider the way the molecules bond to the surface. FIG.\ref{N3_C1s_multi_mono_fit} shows the C 1\textit{s} and Ru 3\textit{d} XPS data for the multilayer (a) and the monolayer (b).

The normalised areas of the peaks in the multilayer should be in the ratio 6:4:2:1 when this ratio represents Pyridine ring C-C:Pyridine ring C-N:Carboxylic acid:Thiocyanate. The actual ratio when normalised to the Thiocyanate is 5.3:3.1:1.8:1.0 for the multilayer and 5.8:3.7:2.2:1 for the monolayer. Hence there is good agreement with expected results. Peak assignment also agrees with previously published data.\cite{Mayor2009a} No change in energy of the C 1\textit{s} peaks related to pyridine and thiocyanate was observed. These parts are in the same chemical environment in both the monolayer and multilayer, which indicates that these parts of the molecule are not involved in bonding to the surface. The carboxylic acid peak is \SI{7.6}{eV} above the Ru 3\textit{d} peak in the case of partial monolayers and this shifts to \SI{7.4}{eV} for the full-monolayer and multilayer coverages. This could be due to the carboxylic acid groups, which are not bound to the surface, being involved in intermolecular hydrogen-bonding at coverages of a monolayer and above.

\begin{figure}[!h]
\centering
\includegraphics[width=8cm]{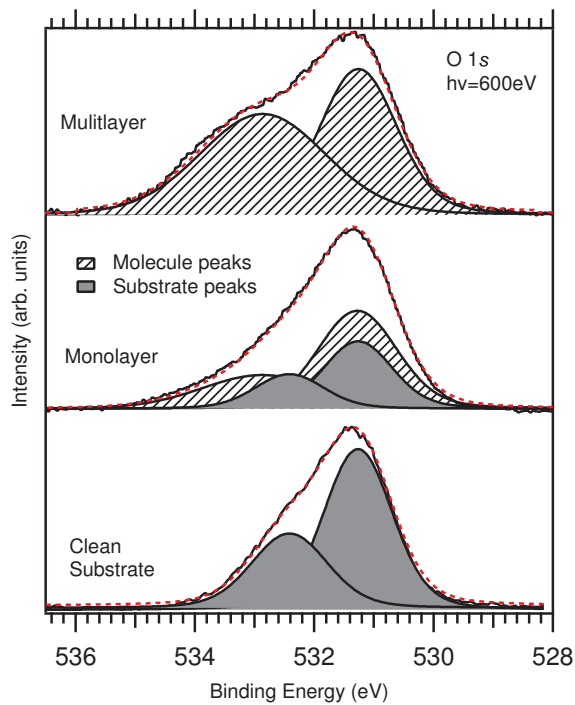}
\caption{O 1\textit{s} Plots normalised to the surface peak, black is background data of the clean surface with identical peak position and shape in the clean surface and monolayer, striped peaks are from the molecule with identical peak position and shape in the multilayer and monolayer. The ratio of the peaks in the monolayer suggest a combination of single and double de-protonation is involved when the molecule bonds to the surface. h$\nu$=\SI{600}{eV}.}
\label{N3_O1s_Mono_Multi}
\end{figure}

Analysis of the O 1\textit{s} spectra should indicate if the carboxylic acid group deprotonates and forms chemical bonds with the surface, XPS O 1\textit{s} data is shown in FIG. \ref{N3_O1s_Mono_Multi}. There are two O 1\textit{s} peaks from the ultra-thin oxide layer with an intensity ratio of 2:1. The smaller peak at 533.0 eV binding energy is assigned to a fraction (8 out of 28) of the O 1\textit{s} surface atoms, while the higher intensity peak at 531.3 eV is due to the remaining surface atoms and the interface layer of oxygen atoms which lie directly on top of AlNi(110).\cite{Martin2011} Two peaks related to the molecule are also observed in the monolayer and multilayer O 1\textit{s} spectra. The peak at \SI{533.0}{eV} binding energy is assigned to the C-OH carbon atoms\cite{Mayor2008a} and thus represents carboxylic acid groups which have not deprotonated. The peak at \SI{531.3}{eV} is assigned to C=O and deprotonated COO-.\cite{Mayor2008a} In the multilayer, the ratio of the two peak areas is 1:1, which is to be expected as we have a 1:1 ratio of C-OH:C=O in the unbound molecule, where the carboxylic acid groups remain protonated.

In order to fit the monolayer data the peak position and shape of the multilayer and substrate peaks were kept the same as the multilayer and surface respectively. Only the intensity was adjusted to give the best fit, which gives a 2:1 ratio of C=O and COO-:C-OH. If both the carboxylic acid groups on one bi-isonotinic acid ligand had de-protonated and bound to the surface, we would expect to see a 3:1 ratio, as previously observed on TiO$_2$.\cite{Mayor2008a} If only one carboxylic acid group had deprotonated we would expect a 5:3 ratio. Hence we have a different bonding environment to those previously observed for N3. Since a 2:1 ratio doesn't fit with the eight oxygen atoms in the N3 molecule we suggest the molecule can take on a range of different bonding geometries on this surface. Some molecules bond to the surface by a single bond via one deprotonation, while others molecules bond via de-protonation of two carboxylic acid groups on the bi-isonicotinic acid ligand. Similar multi-conformational adsorption geometries have recently been observed via low temperature STM of N3 on TiO$_2$.\cite{Kley2014} It is also possible that bonds to the surface can form without deprotonation, thus forming monodentate structures.\cite{Ojamae2006} Aluminum oxide prepared in UHV is prone to hydroxylation even under UHV conditions, so mismatch in O1s peak ratios could be due to presence of OH groups on the alumina surface. 

\begin{figure}[!h]
\centering
\includegraphics[width=8cm]{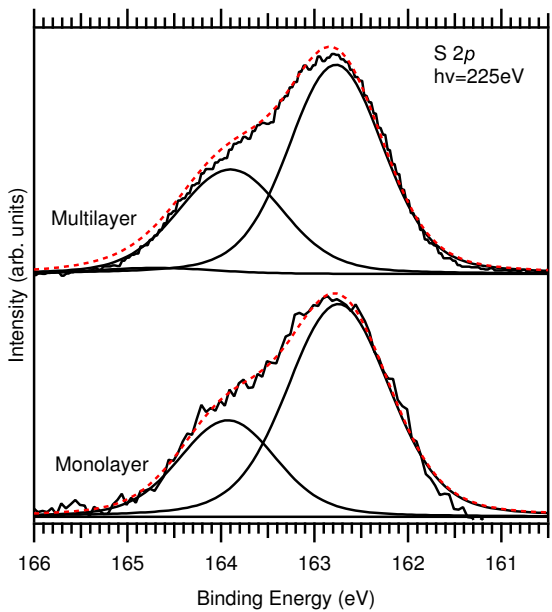}
\caption{S 2\textit{p} XPS showing the spin-orbit split S 2\textit{p}$\frac{1}{2}$ (left) and S 2\textit{p}$\frac{3}{2}$ (right) contributions, exhibiting a single chemical state for the sulphur atoms in the molecule with no binding energy shift at different coverage. Spectra were measured with h$\nu$ = \SI{225}{eV} and normalised to the height of the main peak.}
\label{N3_S2p_Norm_cov}
\end{figure}

Previous results for N3 adsorbed on rutile TiO$_{2}(110)$ exhibited two chemical environments in the S 2\textit{p} region for the monolayer, indicating that the thiocyanate ligands were involved in the bonding to the surface.\cite{Mayor2008a} Here, in contrast, there only one spin-orbit split peak is observed in FIG. \ref{N3_S2p_Norm_cov} for both the monolayer and multilayer, indicating that the thiocyanate group, is not involved in the adsorption bonding.

\begin{figure}[!h]
\centering
\includegraphics[width=8cm]{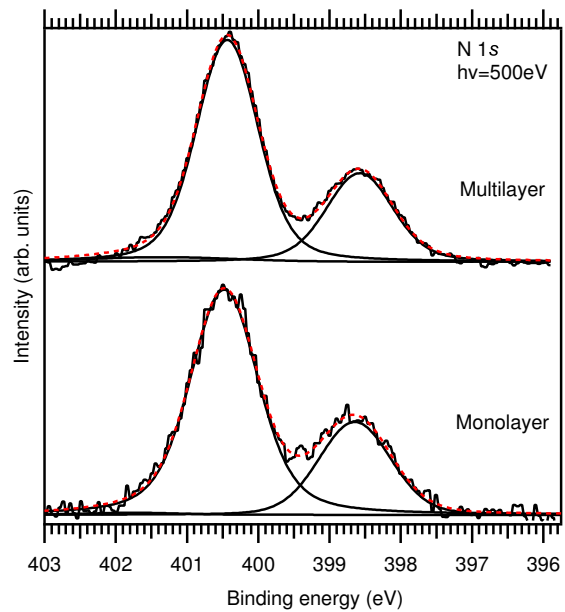}
\caption{N 1\textit{s} XPS measured for multilayer and monolayer of N3, indicating no change in chemical environment between different coverage. The larger peak is due to nitrogen in the bi-isonicotinic acid ligands, while the smaller peak is due to the thiocyanate ligands, with a 2:1 intensity ratio. Spectra were measured with h$\nu$ = \SI{500}{eV} and normalised to the height of the main peak.}
\label{N3_N1s_Multi_Mono}
\end{figure}

XPS measurements of the N 1\textit{s} region, shown in FIG.\ref{N3_N1s_Multi_Mono}, show no change at different coverages, indicating the nitrogen atoms are not involved in bonding to the surface. This provides further evidence that the molecule is intact on the surface.

\subsection{Electronic coupling}

\begin{figure}[!h]
\centering
\includegraphics[width=8cm]{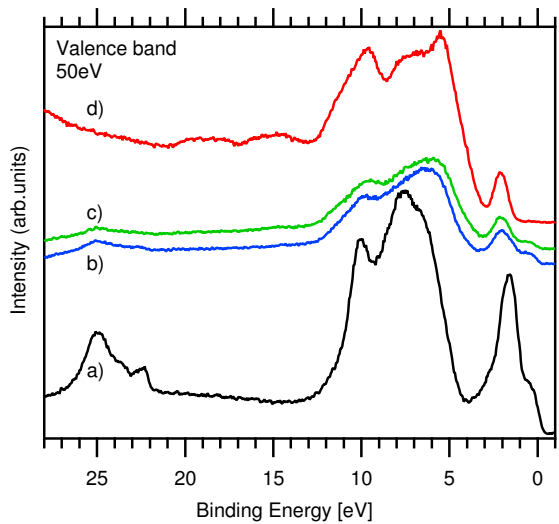}
\caption{Valence band plots at varying coverages, a) clean substrate (black). Higher lines represent increasing coverage. b)  partial monolayer (blue), calculate at 0.3ML c) monolayer (green) 0.85ML d) is a thick film or multilayer  (red) 9ML. All taken at h$\nu$=\SI{50}{eV}.}
\label{N3_VB_Cov_Stack}
\end{figure}

The occupied molecular orbitals and substrate densities of states was measured as a function of surface coverage. The valence band photoemission is shown in FIG.\ref{N3_VB_Cov_Stack} for a multilayer (d), monolayer (c), partial monolayer (b) and the clean oxide surface layer (a). Detailed peak assignments were made on the basis of previously published data for the clean surface.\cite{Andersson1999,Jaeger1991} In these papers it was demonstrated that the peak around \SI{2}{eV} is due to the underlying Ni \textit{d} states in the bulk of the alloy. While the peaks between \SI{5}{eV} and \SI{10}{eV} are due to the oxide layer, which was consistent with previously published angle resolved spectra.\cite{Jaeger1991}

\begin{figure}
\centering
\includegraphics[width=8cm]{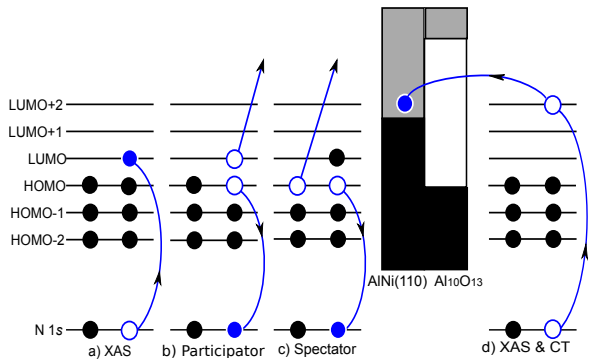}
\caption{Electron excitation and subsequent core-hole induced decay processes: (a) x-ray absorption, resonant core-level excitation into unoccupied bound states; (b) participator decay; (c) spectator decay; and (d) x-ray absorption in the presence of charge transfer from molecular orbital into states near the fermi level of the metal substrate. For the substrate solid colour black white and grey represents occupied states, unoccupied states and band gap respectively.}
\label{Electron_excitation}
\end{figure}

\begin{figure}[!h]
\centering
\includegraphics[width=8cm]{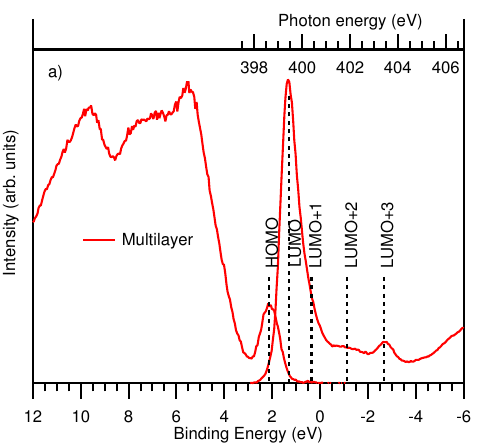}
\includegraphics[width=8cm]{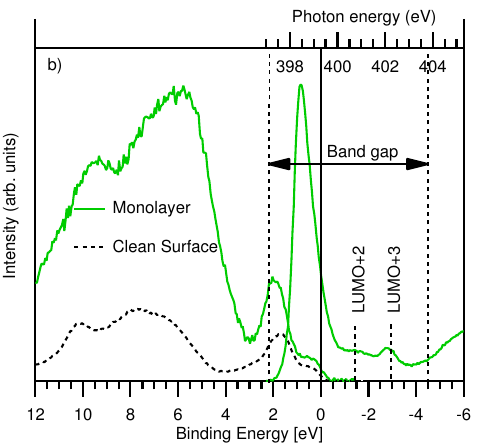}
\caption{Density of States plot constructed from N 1\textit{s} NEXAFS and VB data, (a) multilayer and (b) monolayer coverage with the underlying substrate in black indicating a clear overlap between the unoccupied states of the molecule with the conduction edge of the AlNi(110) substrate.}
\label{DoS_Mono_Blue}
\end{figure}

The unoccupied molecular orbitals can be probed by NEXAFS, in this case at the N 1\textit{s} absorption edge. This process is illustrated for excitation of the core-electron into the LUMO in FIG.\ref{Electron_excitation}(a). If this resonantly excited state overlaps energetically with empty states in the substrate then charge transfer can occur. In the case of a thin oxide film on a metallic substrate there are two relevance conduction bands; that of the oxide and of the metal substrate. The band gap of aluminium oxide at \SI{6.7}{eV}\cite{Andersson1999} is too large for there to be any overlap of the oxide conduction band with the LUMO states of the core-excited molecule. However, if the oxide film is thin enough the excited electron may tunnel through the oxide and into the conduction band of the metal surface as shown in FIG.\ref{Electron_excitation}(d), but only for those states that lie energetically above the Fermi level of the metal surface. To identify the relevant states that can participate in charge transfer from the molecule to the surface for N3 on Al$_{10}$O$_{13}$, the occupied and unoccupied states probed by valence band photoemission and x-ray absorption can be placed on a common binding energy scale using methods described in detail elsewhere.\cite{Schnadt2003}. This data is shown for the N3 monolayer and multilayer in FIG.\ref{DoS_Mono_Blue}.

In FIG.\ref{DoS_Mono_Blue} the N 1\textit{s} (Auger yield) NEXAFS measured across the photon energy range 397-\SI{407}{eV} and placed on the binding energy scale via the N 1\textit{s} binding energy in the pyridine group \SI{400.8}{eV}. The binding energy of the pyridine N 1\textit{s} core level was taken in preference to the N 1\textit{s} binding energy of the thiocynate group, This is based on density functional theory (DFT) calculations\cite{Mayor2008a} showing that the LUMO is located on the bi-isonicotinic acid ligand and the central Ru atom, with no intensity on around the thiocyanate ligand. The photon energy scale from the NEXAFS is also indicated. The HOMO-LUMO gap in the monolayer is 1.5 eV, and 0.8 eV in the multilayer. This is most likely due to surface screening in the monolayer of the excitonic energy shift induced by the core-hole,\cite{Schnadt2005} resulting in the LUMO states being pulled down less energetically in the monolayer than in the multilayer. The band gap of the oxide layer is drawn on as a guide, its location is taken from\cite{Andersson1999} and positioned relative to the Fermi edge. The position of the band gap indicates that charge transfer from the LUMO through to the LUMO+3 states into the oxide layer is not possible, thus any observed charge transfer must be tunneling through the oxide layer and into the AlNi(110) substrate. Data for the monolayer shows that the LUMO+2 and LUMO+3 states of the molecule lie above the Fermi edge and therefore they overlap with the conduction band edge of the underlying AlNi, while the LUMO overlaps occupied states in the surface. This indicates that charge transfer into the substrate from the LUMO+2 and LUMO+3 of the monolayer is possible, but would not be available from the LUMO, and only partly available to the LUMO+1.

\subsection{Charge transfer dynamics}

\begin{figure}
\centering
\includegraphics[width=8cm]{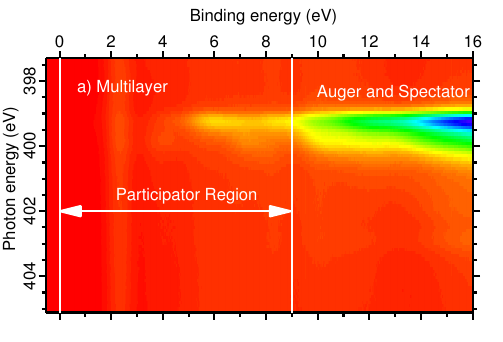}
\includegraphics[width=8cm]{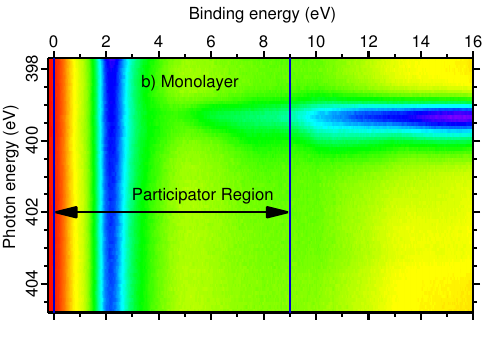}
\caption{N 1\textit{s} RPES multilayer (a) which is the sum of six individual RPES images, four of which are from,\cite{Mayor2008a,Weston2011a} the monolayer (b) is also the sum from six individual RPES spectra. All images were normalised to the beam intensity. The integration window (0-9eV) for the core-hole-clock calculation is also indicated.}
\label{RPES_mono_tot}
\end{figure}

In order to calculate an upper limit on the charge transfer time, we use the core-hole clock implementation of RPES.\cite{Bruhwiler2002} FIG.\ref{RPES_mono_tot} shows 2D RPES datasets for the multilayer (a) and the monolayer (b), each of which have been compiled from six individual data sets. The multilayer data shows an enhancement at \SI{2}{eV} binding energy at the absorption photon energy of the LUMO and LUMO+1, which is attributed to participator decay resulting in resonant photoemission of the HOMO (only weakly visible in FIG.\ref{RPES_mono_tot}). This is described schematically in FIG.\ref{Electron_excitation}(b). Here, the excited electron is emitted in an Auger-like decay process and the final state of the atom is identical to photoemission from the HOMO state. Similar enhancements are also observed for the other occupied molecular orbitals to varying degrees, including a small enhancement at \SI{4}{eV} binding energy and approximately \SI{400}{eV} photon energy (LUMO+1), and also a strong enhancement around \SI{6}{eV} binding energy and \SI{399.2}{eV} photon energy (LUMO). In order to measure the charge transfer time and enable core-hole clock analysis\cite{Bruhwiler2002} this data was integrated over the region 0-\SI{9}{eV} binding energy. In this binding energy window there is some contribution from Auger and spectator decay in the region close to the photon energy of the LUMO, however proportionally this is the same for the monolayer and multilayer. Since this feature tracks out at constant kinetic energy, the contribution will be negligible at the LUMO+2\&3 photon energies, where only the participator channel will be probed.

\begin{figure}[!h]
\centering
\includegraphics[width=8cm]{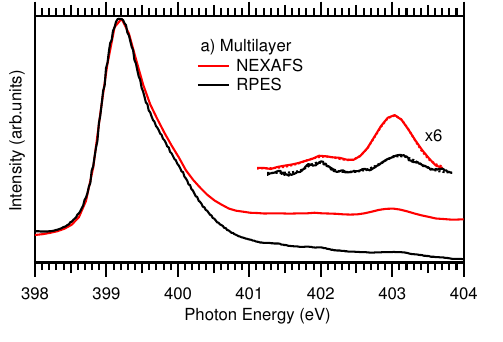}
\includegraphics[width=8cm]{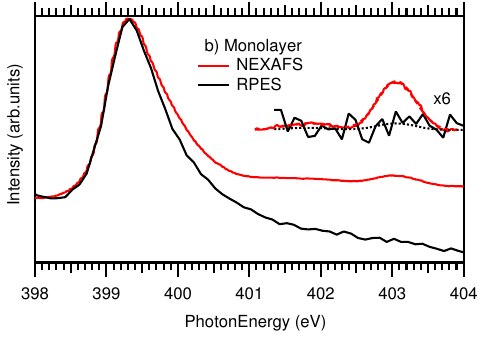}
\caption{N 1\textit{s} RPES and N 1\textit{s} NEXAFS spectra for the N3 multilayer (a) and monolayer (b). The multilayer data are integrations from \SI{0}{eV} to \SI{9}{eV} over all datasets. Also shown are the 6$\times$ magnification of the LUMO+2 and LUMO+3 regions after background removal.}
\label{COC}
\end{figure}

The results of the integration are shown in FIG.\ref{COC} where the RPES and corresponding NEXAFS from multilayer and monolayer data are normalised to the LUMO. The background in the RPES is due to direct photon emission of the valence band, the cross-section for those states decrease with increases increasing photon energy, leading to an increase in the sloping background of the NEXAFS compared to the RPES. The LUMO+1 is too close in energy to the LUMO for these peaks to be separated. Charge transfer from the LUMO into the surface is not possible since the LUMO lies below the fermi level. We also assume that charge transfer from the surface into the LUMO does not occur on the timescale of the core-hole lifetime since this would result in superspectator decay features as previously observed for bi-isonicotinic acid adsorbed on a Au(111) surface.\cite{Taylor2007a} Such features are not observed here. By normalising the data to the LUMO channel for which no charge transfer is allowed for both the monolayer and multilayer, changes in the intensity of the participator channel due to charge transfer out of the LUMO+2 and LUMO+3 states can be probed.

The NEXAFS represents the full intensity of the unoccupied levels, whereas in the case of RPES the unoccupied states may be depleted by charge transfer from the LUMO+2 and LUMO+3 into the substrate, hence in FIG.\ref{COC}  the LUMO+2 and LUMO+3 region of the RPES signal is lower than the NEXAFS. The charge transfer must compete with other decay channels and therefore must be completed within the life-time of the N 1\textit{s} core hole. Hence, the charge transfer time can be measured relative to the lifetime of the core hole. The resonant channels are populated in the multilayer, since the excited electron cannot transfer from the isolated molecule, this represent the maximum intensity resonant channel. The relative depletion of this channel in the monolayer must therefore be due to coupling with the surface. By comparing the relative heights of the NEXAFS to the RPES in both the monolayer and multilayer we can calculate an upper limit for the charge transfer time:\cite{Bruhwiler2002}

\begin{equation}
\tau_{EI}=\tau_{CH}\frac{I_{RPES}^{mono}/I_{NEXAFS}^{mono}}{I_{RPES}^{multi}/I_{NEXAFS}^{multi}-I_{RPES}^{mono}/I_{NEXAFS}^{mono}}
\end{equation}

The $I_{RPES}^{mono}$ and $I_{RPES}^{multi}$ terms represent the intensities of the LUMO+2 and LUMO+3 peaks from the RPES data of the monolayer and multilayer respectively. $I_{RPES}^{multi}$=0.35 while, $I_{RPES}^{mono}$=0.13. The constant $\tau_{CH}$ is the average lifetime of the N 1\textit{s} core hole which is 6.6fs,\cite{Kempgens1996a} hence there is an upper limit to the charge injection time of $6.0\pm$\SI{2.5}{fs}. The charge transfer time for the LUMO+2 and LUMO+3 individually were $6.5\pm$\SI{3.5}{fs} and $6.0\pm$\SI{2.0}{fs} respectively. This compares to upper limits of \SI{4.4}{fs} for N3 on Au(111),\cite{Britton2011} \SI{12}{fs} on TiO$_{2}$,\cite{Weston2011a} and \SI{3}{fs} for bi-isonicotinic acid on TiO$_{2}$.\cite{Schnadt2002} That charge transfer through the ultra-thin aluminium oxide is possible on very short time scales indicates that this is a viable material for incorporation into DSSCs.

\section{Conclusion}

UHV-compatible electrospray deposition has been used to deposit N3 onto an ultra-thin aluminium oxide layer on AlNi(100) \emph{in-situ}. Photoelectron spectroscopy was employed to study the bonding geometry of the dye complex on this surface. We have demonstrated that the thiocyanate ligand is not involved in bonding with the surface. One or both of the carboxylic acid groups on one bi-isonicotinic group will de-protonate giving a chemical bond to the surface,  with the possible addition of monodentate bonds. The energetic alignment of the system was determined by placing the N 1\textit{s} NEXAFS and valence band photoemission onto a common binding energy scale. This indicated that for the monolayer the LUMO is dragged below the Fermi level, and that charge transfer is possible through the oxide layer and into the substrate. Ultra-thin aluminium oxide layer could be a viable material for DSSCs, since it allows the transfer on a time scale of less than $6.0\pm$\SI{2.5}{fs}. Although reduction in recombination effects will not be as significant as for larger aluminium oxide layers, the charge injection from the dye will suffer little suppression and the passivated surface may lead to more stable devices. The next step for this research is to build an ultra-thin oxide layer on TiO$_{2}$ to investigate the charge transfer dynamics.

\section{Acknowledgements}

The research leading to these results has gratefully received funding from the European Community's Seventh Framework Programme (FP7/2007-2013) CALIPSO under grant agreement number 312284, the European Commission through the FP7 Initial Training Network SMALL under joint agreement 238804, the UK Engineering and Physical Sciences Research Council (EPSRC), and Molecularspray Ltd through joint PhD funding. We would like to thanks Miss SE Chandler and Miss S Lally for their invaluable support and help. We are also very grateful to the staff of MAX-lab for their technical assistance, especially Dr Alexei Preobrajenski.

\end{document}